\documentclass[%
reprint,
amsmath,amssymb,
aps,
]{revtex4-2}
\usepackage{graphicx}% Include figure files
\usepackage{dcolumn}% Align table columns on decimal point
\usepackage{bm}% bold math
\usepackage[version=3]{mhchem}
\usepackage{textcomp}

\usepackage{color}
\definecolor{niceblue}{RGB}{55,126,184}
\definecolor{nicered}{RGB}{228,26,28}
\definecolor{drkgreen}{RGB}{0,120,90}
\usepackage[colorlinks,breaklinks,bookmarks=true,citecolor=niceblue,linkcolor=nicered,urlcolor=niceblue]{hyperref}

\usepackage{hyperref}
\begin{document}
	
\preprint{APS/123-QED}

\title{Superconductivity at 12 K in  \ce{La2IOs2}: a 5\textit{d} metal with osmium honeycomb layer}

\author{Hajime Ishikawa}
\email{hishikawa@issp.u-tokyo.ac.jp}
\author{Takeshi Yajima}
\author{Daisuke Nishio-Hamane}
\author{Shusaku Imajo}
\author{Koichi Kindo}
\affiliation{%
	Institute for Solid State Physics, University of Tokyo, Kashiwa, Chiba, 277-8581, Japan 
}%

\author{Mitsuaki Kawamura}
\affiliation{%
	Information Technology Center, University of Tokyo, Bunkyo-ku, Tokyo, 113-8658, Japan 
}%

\begin{abstract}
We discovered superconductivity at $T_{\rm{c}}$ = 12 K in a layered compound \ce{La2IOs2} with osmium honeycomb network. Despite heavy constituent elements unfavorable for phonon mediated mechanism, $T_{\rm{c}}$ is the highest among lanthanoid iodides made of lighter elements such as \ce{La2IRu2} with $T_{\rm{c}}$ = 4.8 K. Electronic anomalies are observed below 60 K similar to those observed in \ce{La2IRu2} below 140 K. \ce{La2IOs2} is a layered 5\textit{d} electron system providing a platform to investigate the interplay between the electronic anomaly, superconductivity, and strong magnetic field.
\end{abstract}

\maketitle

\section{Introduction}

4\textit{d} and 5\textit{d} transition metal compounds have gained considerable interest in condensed matter community. Combination of the strong spin-orbit coupling and Coulomb interactions may stabilize a Mott insulating state with nontrivial ground states such as a quantum spin liquid and multipolar orders \cite{witczak2014correlated}. In metallic compounds, the spin-orbit coupling and electron-electron interactions may cause a Fermi surface instability that gives rise to various electronic orders \cite{fu2015parity}.

Realizations of exotic superconductivity have been discussed in 5\textit{d} transition metal compounds. Carrier doping on the spin-orbit coupled Mott insulator such as layered iridate is a possible avenue \cite{you2012doping, hyart2012competition, watanabe2013monte, meng2014odd}. In SrPtAs with an ordered honeycomb layer made of platinum and arsenide, unusual Cooper pairing state that breaks the time reversal symmetry is proposed \cite{nishikubo2011superconductivity,fischer2014chiral,biswas2013evidence}. Pyrochlore oxide superconductor \ce{Cd2Re2O7} exhibits a phase transition at 200 K below which the space inversion symmetry is broken \cite{hanawa2001superconductivity, hiroi2018pyrochlore}. Odd-parity superconductivity is proposed to apper at the vicinity of the inversion breaking order \cite{kozii2015odd, wang2016topological}.

\begin{figure}[h]
	\includegraphics[width=8.6cm]{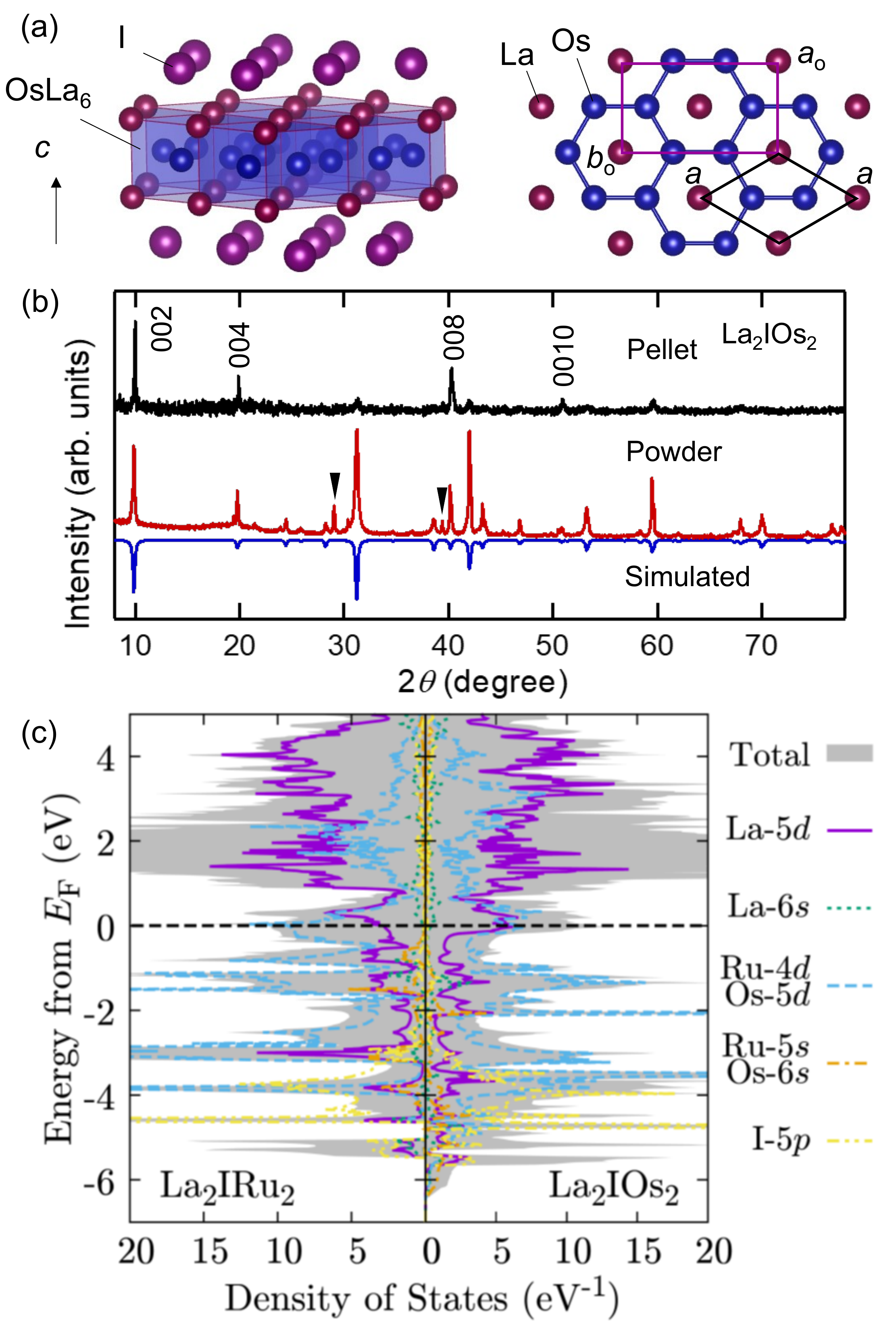}
	\caption{\label{fig:crystal_strct} Crystal structure of \ce{La2IOs2} seen from in-plane direction (left) and from \textit{c}-axis (right) descrived by VESTA \cite{momma2011vesta}. Hexagonal and orthrhombic unit cells are shown by the black and purple lines, respectively. (b) X-ray diffraction patterns of powder (middle) and a pellet (top) of \ce{La2IOs2} compared with the calculated pattern (bottom). Triangles indicate the LaOI impurity phase. (c) Orbital projected density of states calculated on \ce{La2IRu2} (left) and \ce{La2IOs2} (right).}
\end{figure}

In 2019, a peculiar layered superconductor \ce{La2IRu2} with $T_{\rm{c}}$ = 4.8 K is reported \cite{ishikawa2019superconductivity}, which crystallizes in the \ce{Gd2IFe2}-type known as the “intermediate between cluster compounds and intermetallic phases” \cite{ruck1993gd}. The crystal structure features two-dimensional slab made of transition-metal-centered trigonal prism of lanthanoid, where transition metals form the honeycomb network (Fig.~\ref{fig:crystal_strct}(a)). The slabs are separated by iodine anions. The first principles calculation revealed the dominant contributions of Ru-4\textit{d} and La-5\textit{d} orbitals at the Fermi-level. Calculated effective valence for ruthenium was close to -1 based on the Bader charge analysis \cite{baderchargeanalysis}, i.e., the ruthenium is anionic in contrast to cations in honeycomb Mott insulators such as $\alpha$-\ce{RuCl3} \cite{plumb2014alpha}, \ce{Li2RuO3} \cite{miura2007new}, and \ce{SrRu2O6} \cite{hiley2014ruthenium} with valences of +3, +4, and +5, respectively. Superconductivity in \ce{La2IRu2} violates the Pauli-limit expected for the spin-singlet superconductor. Preceding the superconductivity, anomalies are observed in the magnetic susceptibility and the electrical resistivity with small structural changes below 140 K. These observations suggest \ce{La2IRu2} is a layered superconductor made of heavy elements with uncommon electronic states. Little is known about the superconductivity including the universality in the isostructural compounds, the pairing mechanism, and the roles of the spin-orbit coupling of the 4\textit{d} and 5\textit{d} electrons. Here, we sythesized \ce{La2IOs2} \cite{park1997three}, where 5\textit{d}-Os replaces the 4\textit{d}-Ru in \ce{La2IRu2}, and investigated its physical properties. We discovered superconductivity at 12 K preceded by the electronic and structural anomalies below 60 K.

\begin{figure}[h]
	\includegraphics[width=8.6cm]{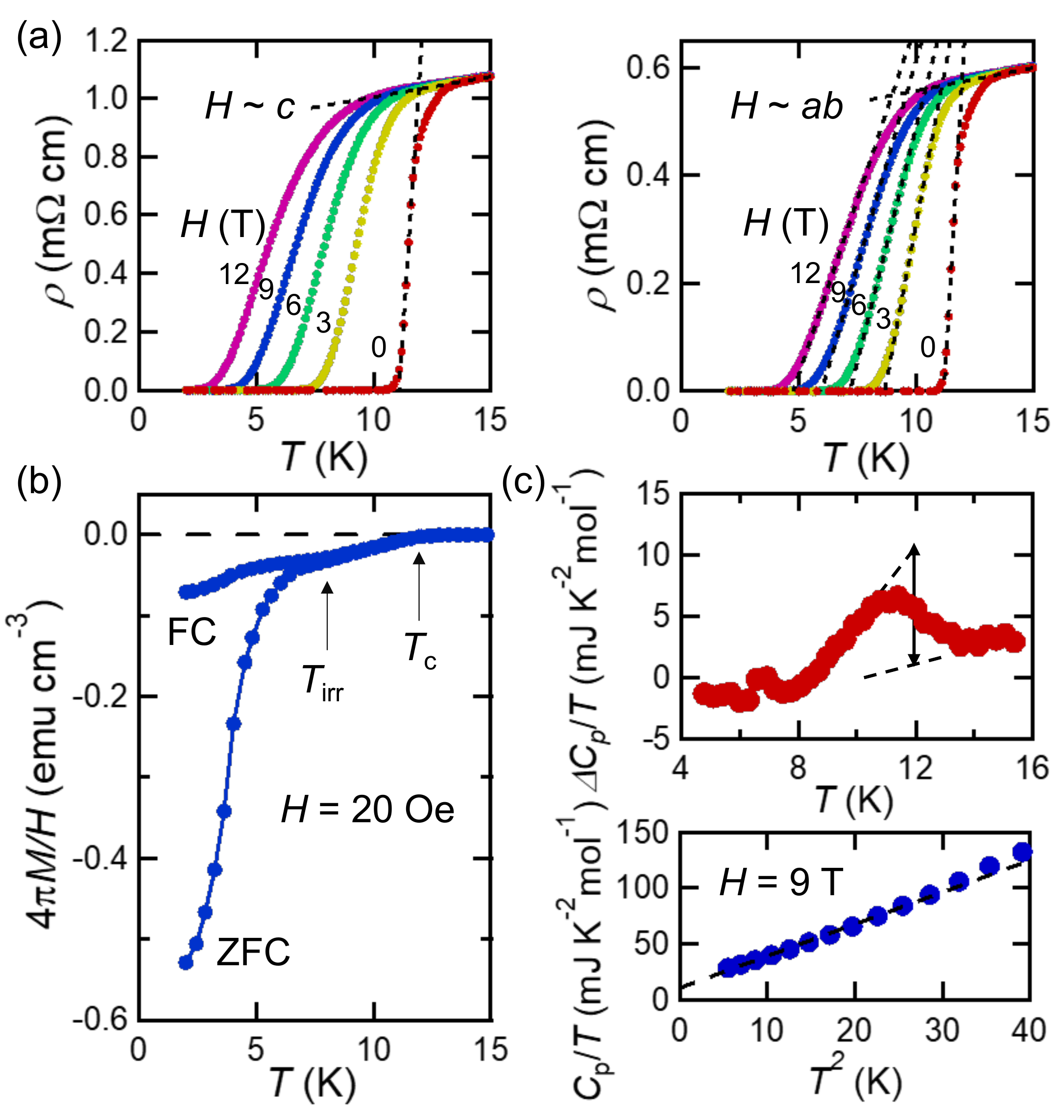}
	\caption{\label{fig:meissner} (a) Temperature dependences of electrical resistivity in magnetic field perpendicular (left) and parallel (right) to the pellet: different pellets from the same powder sample are used. (b) Temperature dependence of 4$\pi$\textit{M}/\textit{H} of the powder sample of \ce{La2IOs2} at \textit{H} = 20 Oe. (c) Difference of the specific heat devided by temperature between the 0 and 9 T data (top): the dashed lines and the arrow are the eye guide. 9 T data is plotted against \textit{T}$^{2}$ (bottom).}
\end{figure}

\section{Sample Preparation}
Powder sample of \ce{La2IOs2} was synthesized by a solid state reaction from the powders of \ce{LaI3}, La (99.9\%), and Os (99.9\%). \ce{LaI3} was synthesized by reacting small pieces of La and iodine (99.999\%) with a molar ratio of 1:4 at 950$^\circ$C for 12 h inside an evacuated quartz glass tube. Excess iodine was removed by sublimation after the reaction. La powder was prepared by filing the rod of La metal. \ce{LaI3}, La, and Os were weighed with the molar ratio of 1:5:6 and mixed in an agate mortar. The powder was sealed inside a tantalum tube and was further sealed inside an evacuated quartz glass tube. The tube was heated at 900 $^\circ$C for 6 days. The reaction yields a dark grey powder. Note that all the experimental procedures were performed inside an argon filled glove box because the starting and target materials are air and moisture sensitive, and Os may produce toxic oxides when exposed to air. To obtain single crystals, similar reaction was performed in more inhomogeneous condition by using small pieces of La instead of the powder as a starting material. The starting materials were sealed as described above and reacted at 950$^\circ$C for 1 day and annealed at 900$^\circ$C for 6 days. The reaction yields inhomogeneous solids including small hexagonal platelet crystals \cite{Supple}. 

Powder x-ray diffraction measurements were performed by a diffractometer with Cu-K$\alpha$ radiation (Smart Lab, Rigaku). The diffraction pattern was analyzed by using the Fullprof software \cite{rodriguez1993recent}. The main peaks observed in the powder pattern shown in the Fig.~\ref{fig:crystal_strct}(b) matches with the simulated pattern of \ce{Gd2IFe2}-type structure with the space group \textit{P}$6_{3}$/\textit{mmc} and lattice constants \textit{a} = 4.2995(1) \AA \ and \textit{c} = 17.953(1) \AA \, at 300 K, which are close to the values in the previous report \cite{park1997three}. Extra peaks can be attributed to the small amount of LaOI impurity phase. LaOI with La$^{3+}$ (4\textit{f}$^{0}$) ion does not possess active electrons and can be negligible in the phsical properties measurements. Chemical analysis of the hexagonal crystals used for the torque measurements were performed by using scanning electron microscopy (SEM, JEOL IT-100) equipped with energy dispersive X-ray spectroscopy (EDS, 15 kV, 0.8 nA), which revealed the chemical composition close to \ce{La2IOs2} \cite{Supple}.

\section{Experimental Methods}
Electrical resistivity measurements were performed by the four-terminal method using the commercial apparatus (PPMS, Quantum Design). Thin plate-shaped pellet is made by pressing the powder sample and used for the measurements. The powder x-ray diffraction pattern for the pellet mounted parallel to the sample holder exhibits predominantly (00\textit{l}) diffraction peaks as in the Fig.~\ref{fig:crystal_strct}(b), indicating sizable preffered orientation along \textit{c}-axis. Au-wires were attached on the pellet by a silver paint. To prevent the degradation of the sample in air at the sample installation for measuremnets, the whole sample was covered by an epoxy resin after the wiring. Magnetization measurements of the powder sample at 2-300 K and at 0-6 T were performed by a SQUID magnetometer (MPMS-XL, Quantum Design). The powder was sealed inside a plastic tube or Al-foil and the corresponding background signal was subtracted. Specific heat measurements were performed by the relaxation method using a commercial apparatus (PPMS, Quantum Design). A small piece of the powder pellet was attached to the sample stage by the apiezon-N grease. Magnetic torque was measured by a microcantilever technique. A hexagonal plate-shaped crystal with approximately 100 $\mu$m size is picked up inside the Ar-filled glove box. The crystal was covered by the apiezon-N grease and attached to the microcantilevers \cite{Supple}, which can be rotated in the magnetic field. The magnetic torque of the crystal was measured as the change in the resistance of the microcantilever.

\section{Calculation Methods}

The electronic states of \ce{La2IOs2} and \ce{La2IRu2} were investigated by the first-principles electronic-structure calculation based on density functional theory. We employed a plane-waves and pseudopotentials-based program package Quantum ESPRESSO~\cite{giannozzi2017advanced}. Perdew-Burke-Ernzerhof's density functional~\cite{perdew1996generalized} based on generalized gradient approximation (GGA) was used. To represent the wavefunction and the atomic potential, we set the cutoff energy to 80 Ry and used fully-relativistic optimized norm-conserving pseudopotential~\cite{scherpelz2016implementation} since we include the spin-orbit interaction in the calculation. The Brillouin-zone integration in the calculation of charge density (density of states) was performed on a 10$\times$10$\times$2 (20$\times$20$\times$4) $\textbf{k}$-point grid by using the optimized tetrahedron method~\cite{kawamura2014improved}. To analyze the contribution from each atom, we computed the projected density of states by projecting each Kohn-Sham state onto atomic orbitals used in the generation of pseudopotentials. Bader charge analysis~\cite{baderchargeanalysis} was performed using a code provided by the Henkelman group~\cite{tang2009grid} together. For this analysis, we employed a projector augmented wave~\cite{blochl1994projector} dataset in PSLibrary~\cite{dal2014pseudopotentials} with a cutoff energy of 75 Ry. The calculated orbital projected density of state is shown in Fig.~\ref{fig:crystal_strct}(c). The band dispersion is shown in the supplemental material \cite{Supple}.

\section{Results} 
Electrical resistivity $\rho$(\textit{T}) measured on the powder pellet is shown in (Fig.~\ref{fig:meissner}(a)). At zero magnetic field, the $\rho$(\textit{T}) decreases linearly below 20 K and deviates downward below 13 K. The zero resistance is observed at 11 K, indicating a superconducting transition. The linear extrapolations of the $\rho$(\textit{T}) above 13 K and the fit in the temperature range where $\rho$ steeply decreases cross at 12 K, providing an estimation of $T_{\rm{c}}$ = 12 K. The decrease in $\rho$ above $T_{\rm{c}}$ may be attributed to the superconducting fluctuations or a certain inhomogenety in the sample.

To check the Meissner effect, we measured magnetization \textit{M} on the powder sample in a weak magnetic field (Fig.~\ref{fig:meissner}(b)). After zero field cooling (ZFC), the large diamagnetic signal is observed. In the magnetic field at \textit{H} = 20 Oe, 4$\pi$\textit{M}/\textit{H}, which corresponds to the superconducting volume fraction, is larger than 50$\%$ at 2 K, suggesting bulk superconductivity. \textit{M}/\textit{H} increases rapidly as increasing temperature up to around 8 K and gradually increases up to around 12 K. On field cooling (FC), the \textit{M}/\textit{H} takes slightly larger values than in the ZFC process down to around 8 K and a clear ZFC-FC hysteresis is observed below 8 K.

At first glance, the \textit{M}/\textit{H} appears that of the powder containing two superconducting phases with $T_{\rm{c}}$ of 8 and 12 K. However, the absence of clear ZFC-FC hysteresis just below $T_{\rm{c}}$ and the appearance of the hysteresis at lower temperature are often observed in layered superconductors with weak vortex pinning including high-$T_{\rm{c}}$ cuprate \cite{muller1987flux} and Li$_{x}$HfNCl \cite{yamanaka1998superconductivity,hotehama2009effect}. The temperature where the hysteresis appears is called irreversibility temperature $T_{\rm{irr}}$, above which vortex are free to move. The behavior of \textit{M}/\textit{H} suggests \ce{La2IOs2} is a superconductor with $T_{\rm{c}}$ = 12 K and $T_{\rm{irr}}$ = 8 K.

\begin{figure}[tb]
	\includegraphics[width=8.6cm]{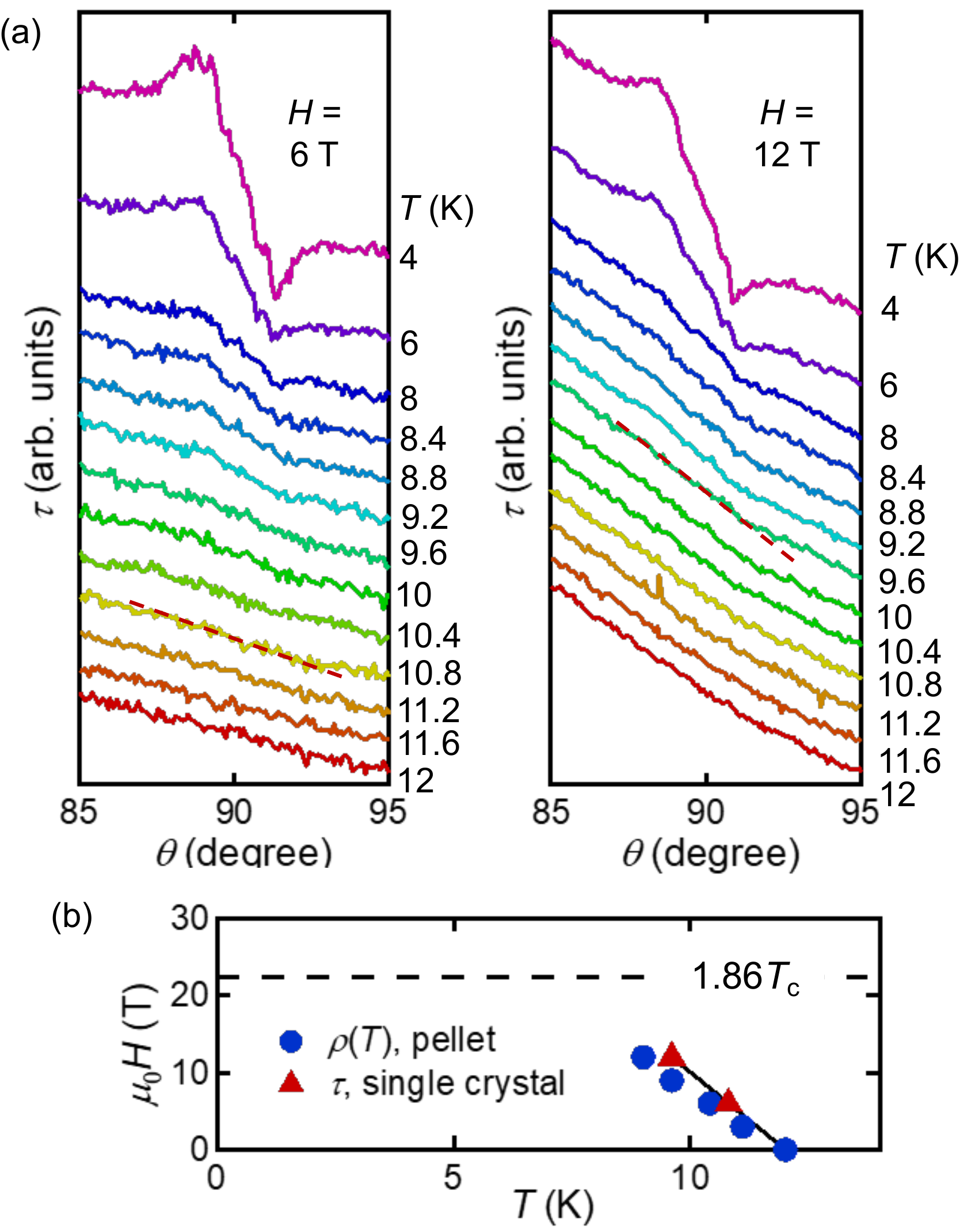}
	\caption{\label{fig:torque} Temperature and magnetic field angle dependence of the magnetic torque of a single crystal of \ce{La2IOs2} at 6 (left) and 12 (right) T. The dashed lines are the eye guide. (b) \textit{H}-\textit{T} phase diagram determined by the resistivity and torque data.}
\end{figure}

To confirm bulk superconductivity at 12 K, we measured the temperature dependence of specific heat \textit{C}$_{p}$ on a thin plate-shaped powder pellet with preffered orientation along \textit{c}-axis (Fig.~\ref{fig:crystal_strct}(c)). Large lattice contribution around $T_{\rm{c}}$ makes the accurate analysis of the electronic contribution challenging. Nevertheless, a peak is observed when the \textit{C}$_{p}$/\textit{T} data at 9 T applied perpendicular to the pellet is subtracted from the 0 T data (Fig.~\ref{fig:meissner}(c)). The peak appears at around 12 K but not around 8 K, supporting the bulk superconductivity at 12 K. Estimated jump of the \textit{C}$_{p}$/\textit{T} at 12 K is approximately 10 mJ/mol-K$^{2}$. \textit{C}$_{p}$/\textit{T} at 9 T is plotted against \textit{T}$^{2}$ and the low-temperature data is linearly fitted assuming \textit{C}$_{p}$ = $\gamma$\textit{T} + $\beta$\textit{T}$^{3}$, where the first and second terms indicate the elecronic and lattice specific heat, respectively. The fit yields $\gamma$ = 11(1) mJ/mol-K$^{2}$ and $\beta$ = 2.85(6) mJ/mol-K$^{4}$. Debye temperature claculated  from $\beta$ is 150 K. Estimated $\Delta$\textit{C}$_{p}$/$\gamma$$T_{\rm{c}}$ is 0.91, which is smaller than 1.43 for the full gap superconductor. Note that the superconductivity is not fully suppressed at 9 T as discussed later and the quantitative discussion is hard based on the present data.

The effect of magnetic field is examined by measuring $\rho$(\textit{T}) in magnetic field applied parallel and perpendicular to the pellet, which can mimic the \textit{H} // \textit{ab} and \textit{c} conditions, respectively (Fig.~\ref{fig:meissner}(a)). The drop of resistivity is suppressed toward low temperatures in both field directions. The effect of magnetic field is weak in \textit{H} parallel to the pellet, which is consistent with the expectation that the orbital pair breaking effect is suppressed when the magnetic field is applied parallel to the superconducting layer. The superconducting transition survives in relatively strong magnetic field at 12 T. 

\begin{figure}[tb]
	\includegraphics[width=8.6cm]{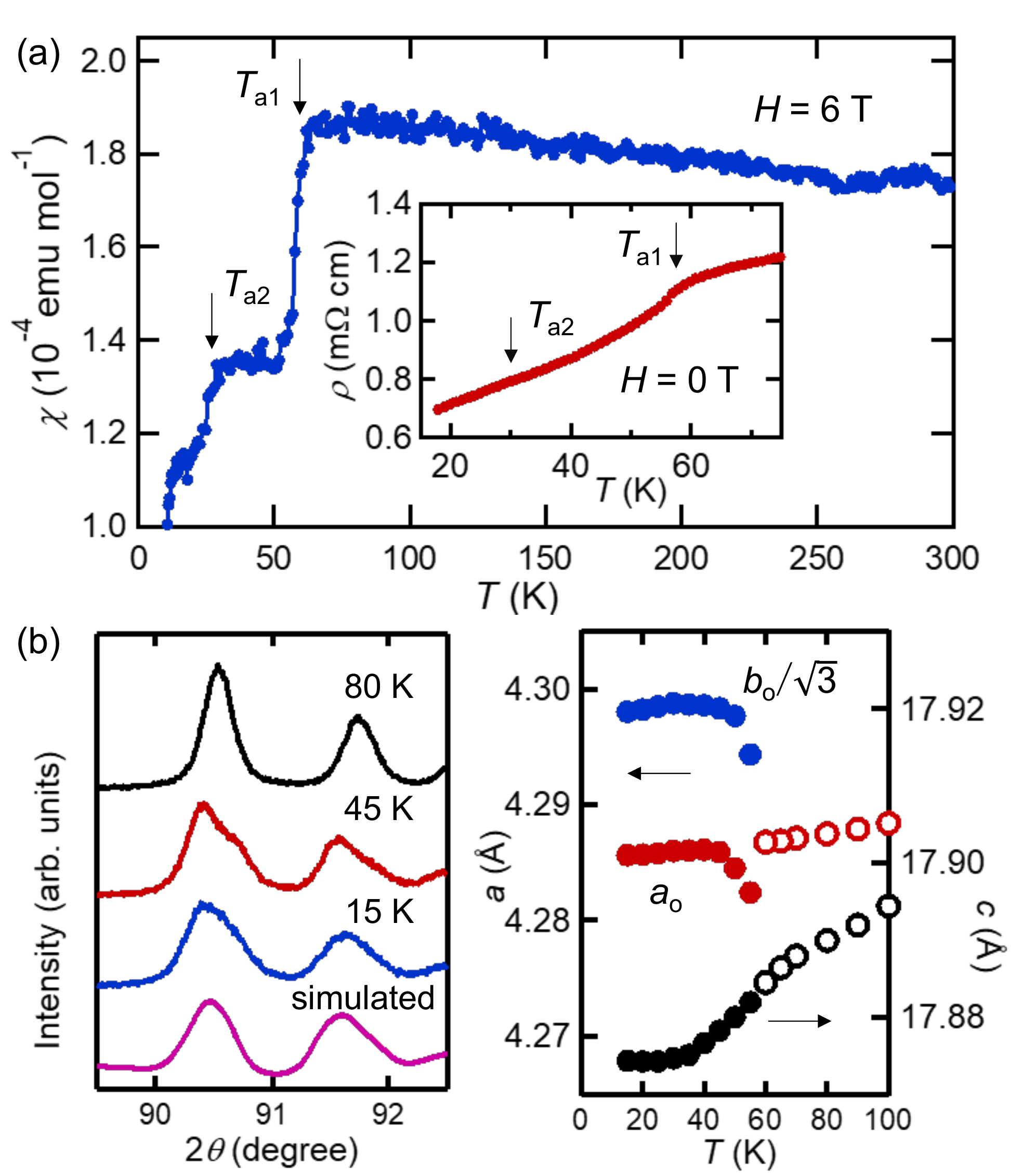}
	\caption{\label{fig:hightemp} (a) Temperature dependence of the magnetic susceptibility at 6 T of the powder sample of \ce{La2IOs2}. Electrical resistivity at 0 T is shown in the inset. (b) Powder x-ray diffraction pattern of \ce{La2IOs2} at selected temperatures compared with the simulation assuming the orthorohmbic unit cell (left). Temperature dependence of the lattice constant of \ce{La2IOs2} (right): values for hexagonal and orthorhombic unit cells are shown by the open and filled plots, respectively.}
\end{figure}

To estimate the upper critical field in \textit{H} // \textit{ab} more precisely, we measured the magnetic torque $\tau$ on a small single crystal (Fig.\ref{fig:torque}(a)). In a layered superconductor, Meissner diamagnetic moment tends to orient perpendicular to the layer because superconducting current may flow more easily within the layer. As $\tau$ is proprtional to \(\boldsymbol{M} \times \boldsymbol{H}\), the torque signal is nearly zero at \textit{H} // \textit{c}, while large signal is observed near \textit{H} // \textit{ab}. In the magnetic field angle dependence of $\tau$, a sharp anomaly is expected across \textit{H} // \textit{ab}, where the $\tau$ changes the sign: the rotation angle $\theta$ = 90$^\circ$ correspond to \textit{H} // \textit{ab} in our data. Indeed, we observed a sharp anomaly at $\theta$ = 90$^\circ$ at 4 K in \textit{H} = 6 and 12 T. At elevated temperatures, the anomaly becomes broader and smaller but visible up to at around 10.8 and 9.6 K at 6 and 12 T, respectively. Note that the magnetic torque of the single crystal is insensitive to the impurities on the surface of the crystal because they typically do not exhibit anisotropy. The observation of the clear torque signal evidences the bulk superconductivity.

\textit{H}-\textit{T} phase diagram in \textit{H} // \textit{ab} is obtained assuming the temperature where the anomaly of $\tau$ is barely visible as $T_{\rm{c}}$. The temperature is slightly higher than $T_{\rm{c}}$ estimated from the linear interpolations of the $\rho$(\textit{T}) data on the powder pellet as described above (Fig.\ref{fig:torque}(a)) likely because the powder averaging is avoided and/or $\tau$ is sensitive to the superconducting fluctuations \cite{tsuchiya2012fluctuating}. The linear fit for the $T_{\rm{c}}$ at 6 and 12 T estimated from the $\tau$ data and $T_{\rm{c}}$ = 12 K at 0 T yields a slope of the \textit{H}-\textit{T} diagram \textit{dH}/\textit{dT} = -5 T/K. The orbital pair breaking field can be estimated by so-called WHH theory as approximately 0.7 $\times$ $|$\textit{dH}$/\textit{dT}$ $|$$_{T=T_{\rm{c}}}$ $\times$ $T_{\rm{c}}$ \cite{werthamer1966temperature}. The estimation yields 42 T, which is larger than the Pauli-limit of 1.86\textit{T}$_{c}$ = 22.3 T. The violation of the Pauli-limit is expected in \ce{La2IOs2} as in \ce{La2IRu2} \cite{ishikawa2019superconductivity}.

In \ce{La2IRu2}, successive anomalies are observed in the magnetic susceptibility and electrical resistivity at around 140 and 85 K \cite{ishikawa2019superconductivity}. In \ce{La2IOs2}, we observed decreases in magnetic susceptibility at 60 and 30 K (Fig.\ref{fig:hightemp}(a)), which bear resembrance with those observed in \ce{La2IRu2}. We call the temperatures of the high and low temperature anomalies as $T_{\rm{a1}}$ and $T_{\rm{a2}}$, respectively, following those in \ce{La2IRu2}. At $T_{\rm{a1}}$, a clear kink is observed in $\rho$(\textit{T}). The anomaly at $T_{\rm{a2}}$ in $\rho$(\textit{T}) is weak, but there appears a weak change of the slope. 

We performed powder X-ray diffraction measurements at low-temperature on \ce{La2IOs2} to examine possible structural changes at $T_{\rm{a1}}$ and $T_{\rm{a2}}$. In the powder pattern, the splitting or broadening of the peaks are observed below 60 K (Fig.\ref{fig:hightemp}(b)), indicating the lowering of the crystal symmetry. The low-temperature data is analyzed assuming the orthorohombic unit cell (Fig.~\ref{fig:crystal_strct}(a)). The \textit{a}-axis length in the orthorhombic unit cell ($a_{\rm{o}}$) is the same with the hexagonal unit cell, while the \textit{b}-axis length ($b_{\rm{o}}$) corresponds to $\sqrt{3}$\textit{a} of the hexagonal unit cell. The temperature dependence of the refined lattice constants are plotted in Fig.\ref{fig:hightemp}(b): $b_{\rm{o}}$/$\sqrt{3}$ is plotted to compare with \textit{a} in the hexagonal cell. The average of the $a_{\rm{o}}$ and $b_{\rm{o}}$/$\sqrt{3}$ increases below 60 K, while the \textit{c}-axis shrinks. In the case of \ce{La2IRu2}, \textit{a}-axis length decreases but the \textit{c}-axis length slightly increases below $T_{\rm{a1}}$ \cite{ishikawa2019superconductivity}. $a_{\rm{o}}$ and $b_{\rm{o}}$ in \ce{La2IOs2} slightly decreases below 30 K, while the changes are small.

\section{Discussion}
\textit{T}$_{c}$ = 12 K in \ce{La2IOs2} is the highest among layered lanthanoid iodides with related crystal structures. Thus far observed highest \textit{T}$_{c}$ was 10 K in the iodide carbide \ce{Y2I2C2} \cite{henn1996bulk}. In \ce{La2I2C2}, superconductivity is observed below 2 K \cite{ahn2000superconductivity}. In \ce{La2TeI2}, where Te replaces the \ce{C2} units, superconductivity is not reported \cite{ryazanov2006la2tei2}. Suppression of \textit{T}$_{c}$ on substituting \ce{Y2I2C2} by heavier elements appears compatible with a phonon mediated superconductivity, where lighter elements that enhance the phonon frequencies are favored. We expect lower \textit{T}$_{c}$ in heavier \ce{La2IOs2} than in \ce{La2IRu2}: the Debye temperatures estimated from the specific heat measurements are 150 and 162 K in \ce{La2IOs2} and \ce{La2IRu2}, respectively. Contrary to the expectation, \textit{T}$_{c}$ of \ce{La2IOs2} is 2.5 times higher than \ce{La2IRu2}. The electronic specific heat coefficient $\gamma$ is estimated as 11 and 13 mJ/mol-K$^{2}$ on \ce{La2IOs2} and \ce{La2IRu2}, respectively. Smaller $\gamma$ in \ce{La2IOs2} suggests the smaller density of states and is again opposed to the higher \textit{T}$_{c}$.

\ce{La2IOs2} and \ce{La2IRu2} are distinguished from the \ce{Y2I2C2} family by the presence of the high temperature anomalies preceding the superconductivity at $T_{\rm{a1}}$ and $T_{\rm{a2}}$, where similar decreaeses in the magnetic susceptibility are observed. The two compounds are made of the same group elements with similar effective valences verified by the Bader charge calculation: +1.41 (+1.37) for La, -0.80 (-0.79) for I, and -1.01 (-0.98) for Os (Ru) in \ce{La2IOs2} (\ce{La2IRu2}). It would be reasonable to consider the anomalies in the two compounds share the common origin. Enhancement of \textit{T}$_{c}$ on suppression of an electronic order is observed in various layered superconductors including cuprates, iron pnictides, and charge-density-wave systems \cite{keimer2015quantum, si2016high, berthier1976evidence, morosan2006superconductivity, chen2021double}. The relationship between the high-temperature anomalies and \textit{T}$_{c}$ in \ce{La2IOs2} and \ce{La2IRu2} follows this trend, pointing to an interplay between the high temperature anomalies and the superconductivity.

The drop of magnetic susceptibility at $T_{\rm{a1}}$ and $T_{\rm{a2}}$ should indicate the decrease in the density of states at the Fermi energy or an antiferromagnetic order. Antiferromagnetic order is not compatible with the results of nuclear magnetic resonance experiments on \ce{La2IRu2} \cite{ishikawa2019superconductivity}. More delocalized 5\textit{d} orbitals in \ce{La2IOs2} should further destabilize the magnetism \cite{Supple}. A charge-density-wave order can be compatible with the observations. The decrease in resistivity with a kink is similar to those observed at the charge-density-wave order in layered superconductors \cite{berthier1976evidence, morosan2006superconductivity, chen2021double}. Note that there is one transition above the superconductivity in these charge-density-wave systems in contrast to the two transitions observed in \ce{La2IOs2} and \ce{La2IRu2}.

We propose another possiblity by pointing out the striking similarities between \ce{La2IOs2} and 5\textit{d} pyrochlore superconductor \ce{Cd2Re2O7} \cite{hanawa2001superconductivity}. \ce{Cd2Re2O7} exhibits successive phase transitions at 200 and 120 K acompanying the decreases in magnetic susceptibility and resistivity similar to \ce{La2IOs2}. Suppression of the transitions by applying pressure raises the \textit{T}$_{c}$ and enhances the upper critical field above the Pauli-limit \cite{c2011superconductivity}. \ce{Cd2Re2O7} gains attention as the spin-orbit coupled metal, where an interplay between an electronic order and the odd-parity superconductivity is proposed \cite{fu2015parity,hiroi2018pyrochlore,kozii2015odd,wang2016topological}. As seen in the orbital projected density of states (Fig.~\ref{fig:crystal_strct}(c)), the bands around the Fermi-enerygy are dominated by the Os-5\textit{d} and La-5\textit{d} orbitals, where strong effect of spin-orbit coupling is expected. Examining whether the high-temperature anomalies and the superconductivity with high upper critical field are relevant to those discussed in the spin-orbit coupled system would be an intriguing problem.

In addition to the suppressed high temperature anomalies, the difference in the lattice distortions at low temperature may play a role on the enhancement of \textit{T}$_{c}$ in \ce{La2IOs2} compared to \ce{La2IRu2}. The single crystal structural analysis on \ce{La2IRu2} points to the small displacement of Ru atoms along \textit{c}-axis without clear symmetry lowering \cite{ishikawa2019superconductivity}. On the other hand, the powder x-ray diffraction on \ce{La2IOs2} suggests the in-plane distortions. The different lattice distortions observed in \ce{La2IOs2} and \ce{La2IRu2} may indicate different ordering patterns are energetically degenerate. The band structure calculations on the high temperature hexagonal structure revealed the presence of several bands crossing the Fermi energy in both compounds \cite{Supple}. The multiband superconductor may exhibit enhanced \textit{T}$_{c}$ \cite{zehetmayer2013review} as extensively investigated in \ce{MgB2} with \textit{T}$_{c}$ = 39 K \cite{mazin2003electronic} and high upper critical field violating the Pauli-limit as in the Chevrel phases with \textit{T}$_{c}$ $\sim$ 15 K \cite{petrovic2011multiband}. The band structure of \ce{La2IOs2} and \ce{La2IRu2} should be modified in different ways at low temperature. A certain multiband effect that enhances \textit{T}$_{c}$ may exist in the low-temperature band structure of \ce{La2IOs2}.

\section{Conclusion and Perspective}
We discovered superconductivity at \textit{T}$_{c}$ = 12 K in a layered compound \ce{La2IOs2}. Successive anomalies are observed in the electronic properties below 60 K accompanying structural changes. The superconductivity is suppressed only by a few kelvins in magnetic field at 12 T. \ce{La2IOs2} is a 5\textit{d} electron system that allows us to investigate the interplay between electronic anomalies, superconductivity, and high magnetic field in a layered structure featuring honeycomb lattice. Further investigations to clarify the nature of the high temperature anomalies and its relationship between the superconductivity are necessary. If the corresponding order promotes the superconductivity, higher \textit{T}$_{c}$ can be obtained around the quantum critical point, providing so-called superconducting dome in the phase diagram. Examining the evolution of $T_{\rm{a1}}$, $T_{\rm{a2}}$, and \textit{T}$_{c}$ against a certain controll parameter by performing doping or high pressure experiments are desired. Uncovering the \textit{H}-\textit{T} phase diagram at high magnetic field region would be informative for discussing the multiband effect.

\begin{acknowledgments}
We would like to thank Prof. Zenji Hiroi and Prof. Yoshihiko Okamoto for the help in the use of osmium and fruitful discussions. HI is supported by JSPS KAKENHI Grants No. JP22H04467 and No. JP22K13996.
\end{acknowledgments}

\bibliography{LOI}

\begin{thebibliography}{47}%
\makeatletter
\providecommand \@ifxundefined [1]{%
 \@ifx{#1\undefined}
}%
\providecommand \@ifnum [1]{%
 \ifnum #1\expandafter \@firstoftwo
 \else \expandafter \@secondoftwo
 \fi
}%
\providecommand \@ifx [1]{%
 \ifx #1\expandafter \@firstoftwo
 \else \expandafter \@secondoftwo
 \fi
}%
\providecommand \natexlab [1]{#1}%
\providecommand \enquote  [1]{``#1''}%
\providecommand \bibnamefont  [1]{#1}%
\providecommand \bibfnamefont [1]{#1}%
\providecommand \citenamefont [1]{#1}%
\providecommand \href@noop [0]{\@secondoftwo}%
\providecommand \href [0]{\begingroup \@sanitize@url \@href}%
\providecommand \@href[1]{\@@startlink{#1}\@@href}%
\providecommand \@@href[1]{\endgroup#1\@@endlink}%
\providecommand \@sanitize@url [0]{\catcode `\\12\catcode `\$12\catcode
  `\&12\catcode `\#12\catcode `\^12\catcode `\_12\catcode `\%12\relax}%
\providecommand \@@startlink[1]{}%
\providecommand \@@endlink[0]{}%
\providecommand \url  [0]{\begingroup\@sanitize@url \@url }%
\providecommand \@url [1]{\endgroup\@href {#1}{\urlprefix }}%
\providecommand \urlprefix  [0]{URL }%
\providecommand \Eprint [0]{\href }%
\providecommand \doibase [0]{https://doi.org/}%
\providecommand \selectlanguage [0]{\@gobble}%
\providecommand \bibinfo  [0]{\@secondoftwo}%
\providecommand \bibfield  [0]{\@secondoftwo}%
\providecommand \translation [1]{[#1]}%
\providecommand \BibitemOpen [0]{}%
\providecommand \bibitemStop [0]{}%
\providecommand \bibitemNoStop [0]{.\EOS\space}%
\providecommand \EOS [0]{\spacefactor3000\relax}%
\providecommand \BibitemShut  [1]{\csname bibitem#1\endcsname}%
\let\auto@bib@innerbib\@empty
%</preamble>
\bibitem [{\citenamefont {Witczak-Krempa}\ \emph {et~al.}(2014)\citenamefont
  {Witczak-Krempa}, \citenamefont {Chen}, \citenamefont {Kim},\ and\
  \citenamefont {Balents}}]{witczak2014correlated}%
  \BibitemOpen
  \bibfield  {author} {\bibinfo {author} {\bibfnamefont {W.}~\bibnamefont
  {Witczak-Krempa}}, \bibinfo {author} {\bibfnamefont {G.}~\bibnamefont
  {Chen}}, \bibinfo {author} {\bibfnamefont {Y.~B.}\ \bibnamefont {Kim}},\ and\
  \bibinfo {author} {\bibfnamefont {L.}~\bibnamefont {Balents}},\ }\href@noop
  {} {\bibfield  {journal} {\bibinfo  {journal} {Annu. Rev. Condens. Matter
  Phys.}\ }\textbf {\bibinfo {volume} {5}},\ \bibinfo {pages} {57} (\bibinfo
  {year} {2014})}\BibitemShut {NoStop}%
\bibitem [{\citenamefont {Fu}(2015)}]{fu2015parity}%
  \BibitemOpen
  \bibfield  {author} {\bibinfo {author} {\bibfnamefont {L.}~\bibnamefont
  {Fu}},\ }\href@noop {} {\bibfield  {journal} {\bibinfo  {journal} {Phys. Rev.
  Lett.}\ }\textbf {\bibinfo {volume} {115}},\ \bibinfo {pages} {026401}
  (\bibinfo {year} {2015})}\BibitemShut {NoStop}%
\bibitem [{\citenamefont {You}\ \emph {et~al.}(2012)\citenamefont {You},
  \citenamefont {Kimchi},\ and\ \citenamefont {Vishwanath}}]{you2012doping}%
  \BibitemOpen
  \bibfield  {author} {\bibinfo {author} {\bibfnamefont {Y.-Z.}\ \bibnamefont
  {You}}, \bibinfo {author} {\bibfnamefont {I.}~\bibnamefont {Kimchi}},\ and\
  \bibinfo {author} {\bibfnamefont {A.}~\bibnamefont {Vishwanath}},\
  }\href@noop {} {\bibfield  {journal} {\bibinfo  {journal} {Phys. Rev. B}\
  }\textbf {\bibinfo {volume} {86}},\ \bibinfo {pages} {085145} (\bibinfo
  {year} {2012})}\BibitemShut {NoStop}%
\bibitem [{\citenamefont {Hyart}\ \emph {et~al.}(2012)\citenamefont {Hyart},
  \citenamefont {Wright}, \citenamefont {Khaliullin},\ and\ \citenamefont
  {Rosenow}}]{hyart2012competition}%
  \BibitemOpen
  \bibfield  {author} {\bibinfo {author} {\bibfnamefont {T.}~\bibnamefont
  {Hyart}}, \bibinfo {author} {\bibfnamefont {A.~R.}\ \bibnamefont {Wright}},
  \bibinfo {author} {\bibfnamefont {G.}~\bibnamefont {Khaliullin}},\ and\
  \bibinfo {author} {\bibfnamefont {B.}~\bibnamefont {Rosenow}},\ }\href@noop
  {} {\bibfield  {journal} {\bibinfo  {journal} {Phys. Rev. B}\ }\textbf
  {\bibinfo {volume} {85}},\ \bibinfo {pages} {140510} (\bibinfo {year}
  {2012})}\BibitemShut {NoStop}%
\bibitem [{\citenamefont {Watanabe}\ \emph {et~al.}(2013)\citenamefont
  {Watanabe}, \citenamefont {Shirakawa},\ and\ \citenamefont
  {Yunoki}}]{watanabe2013monte}%
  \BibitemOpen
  \bibfield  {author} {\bibinfo {author} {\bibfnamefont {H.}~\bibnamefont
  {Watanabe}}, \bibinfo {author} {\bibfnamefont {T.}~\bibnamefont
  {Shirakawa}},\ and\ \bibinfo {author} {\bibfnamefont {S.}~\bibnamefont
  {Yunoki}},\ }\href@noop {} {\bibfield  {journal} {\bibinfo  {journal} {Phys.
  Rev. Lett.}\ }\textbf {\bibinfo {volume} {110}},\ \bibinfo {pages} {027002}
  (\bibinfo {year} {2013})}\BibitemShut {NoStop}%
\bibitem [{\citenamefont {Meng}\ \emph {et~al.}(2014)\citenamefont {Meng},
  \citenamefont {Kim},\ and\ \citenamefont {Kee}}]{meng2014odd}%
  \BibitemOpen
  \bibfield  {author} {\bibinfo {author} {\bibfnamefont {Z.~Y.}\ \bibnamefont
  {Meng}}, \bibinfo {author} {\bibfnamefont {Y.~B.}\ \bibnamefont {Kim}},\ and\
  \bibinfo {author} {\bibfnamefont {H.-Y.}\ \bibnamefont {Kee}},\ }\href@noop
  {} {\bibfield  {journal} {\bibinfo  {journal} {Phys. Rev. Lett.}\ }\textbf
  {\bibinfo {volume} {113}},\ \bibinfo {pages} {177003} (\bibinfo {year}
  {2014})}\BibitemShut {NoStop}%
\bibitem [{\citenamefont {Nishikubo}\ \emph {et~al.}(2011)\citenamefont
  {Nishikubo}, \citenamefont {Kudo},\ and\ \citenamefont
  {Nohara}}]{nishikubo2011superconductivity}%
  \BibitemOpen
  \bibfield  {author} {\bibinfo {author} {\bibfnamefont {Y.}~\bibnamefont
  {Nishikubo}}, \bibinfo {author} {\bibfnamefont {K.}~\bibnamefont {Kudo}},\
  and\ \bibinfo {author} {\bibfnamefont {M.}~\bibnamefont {Nohara}},\
  }\href@noop {} {\bibfield  {journal} {\bibinfo  {journal} {J. Phys. Soc.
  Jpn.}\ }\textbf {\bibinfo {volume} {80}},\ \bibinfo {pages} {055002}
  (\bibinfo {year} {2011})}\BibitemShut {NoStop}%
\bibitem [{\citenamefont {Fischer}\ \emph {et~al.}(2014)\citenamefont
  {Fischer}, \citenamefont {Neupert}, \citenamefont {Platt}, \citenamefont
  {Schnyder}, \citenamefont {Hanke}, \citenamefont {Goryo}, \citenamefont
  {Thomale},\ and\ \citenamefont {Sigrist}}]{fischer2014chiral}%
  \BibitemOpen
  \bibfield  {author} {\bibinfo {author} {\bibfnamefont {M.~H.}\ \bibnamefont
  {Fischer}}, \bibinfo {author} {\bibfnamefont {T.}~\bibnamefont {Neupert}},
  \bibinfo {author} {\bibfnamefont {C.}~\bibnamefont {Platt}}, \bibinfo
  {author} {\bibfnamefont {A.~P.}\ \bibnamefont {Schnyder}}, \bibinfo {author}
  {\bibfnamefont {W.}~\bibnamefont {Hanke}}, \bibinfo {author} {\bibfnamefont
  {J.}~\bibnamefont {Goryo}}, \bibinfo {author} {\bibfnamefont
  {R.}~\bibnamefont {Thomale}},\ and\ \bibinfo {author} {\bibfnamefont
  {M.}~\bibnamefont {Sigrist}},\ }\href@noop {} {\bibfield  {journal} {\bibinfo
   {journal} {Phys. Rev. B}\ }\textbf {\bibinfo {volume} {89}},\ \bibinfo
  {pages} {020509} (\bibinfo {year} {2014})}\BibitemShut {NoStop}%
\bibitem [{\citenamefont {Biswas}\ \emph {et~al.}(2013)\citenamefont {Biswas},
  \citenamefont {Luetkens}, \citenamefont {Neupert}, \citenamefont
  {St{\"u}rzer}, \citenamefont {Baines}, \citenamefont {Pascua}, \citenamefont
  {Schnyder}, \citenamefont {Fischer}, \citenamefont {Goryo}, \citenamefont
  {Lees} \emph {et~al.}}]{biswas2013evidence}%
  \BibitemOpen
  \bibfield  {author} {\bibinfo {author} {\bibfnamefont {P.~K.}\ \bibnamefont
  {Biswas}}, \bibinfo {author} {\bibfnamefont {H.}~\bibnamefont {Luetkens}},
  \bibinfo {author} {\bibfnamefont {T.}~\bibnamefont {Neupert}}, \bibinfo
  {author} {\bibfnamefont {T.}~\bibnamefont {St{\"u}rzer}}, \bibinfo {author}
  {\bibfnamefont {C.}~\bibnamefont {Baines}}, \bibinfo {author} {\bibfnamefont
  {G.}~\bibnamefont {Pascua}}, \bibinfo {author} {\bibfnamefont {A.~P.}\
  \bibnamefont {Schnyder}}, \bibinfo {author} {\bibfnamefont {M.~H.}\
  \bibnamefont {Fischer}}, \bibinfo {author} {\bibfnamefont {J.}~\bibnamefont
  {Goryo}}, \bibinfo {author} {\bibfnamefont {M.~R.}\ \bibnamefont {Lees}},
  \emph {et~al.},\ }\href@noop {} {\bibfield  {journal} {\bibinfo  {journal}
  {Phys. Rev. B}\ }\textbf {\bibinfo {volume} {87}},\ \bibinfo {pages} {180503}
  (\bibinfo {year} {2013})}\BibitemShut {NoStop}%
\bibitem [{\citenamefont {Hanawa}\ \emph {et~al.}(2001)\citenamefont {Hanawa},
  \citenamefont {Muraoka}, \citenamefont {Tayama}, \citenamefont {Sakakibara},
  \citenamefont {Yamaura},\ and\ \citenamefont
  {Hiroi}}]{hanawa2001superconductivity}%
  \BibitemOpen
  \bibfield  {author} {\bibinfo {author} {\bibfnamefont {M.}~\bibnamefont
  {Hanawa}}, \bibinfo {author} {\bibfnamefont {Y.}~\bibnamefont {Muraoka}},
  \bibinfo {author} {\bibfnamefont {T.}~\bibnamefont {Tayama}}, \bibinfo
  {author} {\bibfnamefont {T.}~\bibnamefont {Sakakibara}}, \bibinfo {author}
  {\bibfnamefont {J.}~\bibnamefont {Yamaura}},\ and\ \bibinfo {author}
  {\bibfnamefont {Z.}~\bibnamefont {Hiroi}},\ }\href@noop {} {\bibfield
  {journal} {\bibinfo  {journal} {Phys. Rev. Lett.}\ }\textbf {\bibinfo
  {volume} {87}},\ \bibinfo {pages} {187001} (\bibinfo {year}
  {2001})}\BibitemShut {NoStop}%
\bibitem [{\citenamefont {Hiroi}\ \emph {et~al.}(2018)\citenamefont {Hiroi},
  \citenamefont {Yamaura}, \citenamefont {Kobayashi}, \citenamefont
  {Matsubayashi},\ and\ \citenamefont {Hirai}}]{hiroi2018pyrochlore}%
  \BibitemOpen
  \bibfield  {author} {\bibinfo {author} {\bibfnamefont {Z.}~\bibnamefont
  {Hiroi}}, \bibinfo {author} {\bibfnamefont {J.-i.}\ \bibnamefont {Yamaura}},
  \bibinfo {author} {\bibfnamefont {T.~C.}\ \bibnamefont {Kobayashi}}, \bibinfo
  {author} {\bibfnamefont {Y.}~\bibnamefont {Matsubayashi}},\ and\ \bibinfo
  {author} {\bibfnamefont {D.}~\bibnamefont {Hirai}},\ }\href@noop {}
  {\bibfield  {journal} {\bibinfo  {journal} {J. Phys. Soc. Jpn.}\ }\textbf
  {\bibinfo {volume} {87}},\ \bibinfo {pages} {024702} (\bibinfo {year}
  {2018})}\BibitemShut {NoStop}%
\bibitem [{\citenamefont {Kozii}\ and\ \citenamefont
  {Fu}(2015)}]{kozii2015odd}%
  \BibitemOpen
  \bibfield  {author} {\bibinfo {author} {\bibfnamefont {V.}~\bibnamefont
  {Kozii}}\ and\ \bibinfo {author} {\bibfnamefont {L.}~\bibnamefont {Fu}},\
  }\href@noop {} {\bibfield  {journal} {\bibinfo  {journal} {Phys. Rev. Lett.}\
  }\textbf {\bibinfo {volume} {115}},\ \bibinfo {pages} {207002} (\bibinfo
  {year} {2015})}\BibitemShut {NoStop}%
\bibitem [{\citenamefont {Wang}\ \emph {et~al.}(2016)\citenamefont {Wang},
  \citenamefont {Cho}, \citenamefont {Hughes},\ and\ \citenamefont
  {Fradkin}}]{wang2016topological}%
  \BibitemOpen
  \bibfield  {author} {\bibinfo {author} {\bibfnamefont {Y.}~\bibnamefont
  {Wang}}, \bibinfo {author} {\bibfnamefont {G.~Y.}\ \bibnamefont {Cho}},
  \bibinfo {author} {\bibfnamefont {T.~L.}\ \bibnamefont {Hughes}},\ and\
  \bibinfo {author} {\bibfnamefont {E.}~\bibnamefont {Fradkin}},\ }\href@noop
  {} {\bibfield  {journal} {\bibinfo  {journal} {Phys. Rev. B}\ }\textbf
  {\bibinfo {volume} {93}},\ \bibinfo {pages} {134512} (\bibinfo {year}
  {2016})}\BibitemShut {NoStop}%
\bibitem [{\citenamefont {Ishikawa}\ \emph {et~al.}(2019)\citenamefont
  {Ishikawa}, \citenamefont {Wedig}, \citenamefont {Nuss}, \citenamefont
  {Kremer}, \citenamefont {Dinnebier}, \citenamefont {Blankenhorn},
  \citenamefont {Pakdaman}, \citenamefont {Matsumoto}, \citenamefont
  {Takayama}, \citenamefont {Kitagawa} \emph
  {et~al.}}]{ishikawa2019superconductivity}%
  \BibitemOpen
  \bibfield  {author} {\bibinfo {author} {\bibfnamefont {H.}~\bibnamefont
  {Ishikawa}}, \bibinfo {author} {\bibfnamefont {U.}~\bibnamefont {Wedig}},
  \bibinfo {author} {\bibfnamefont {J.}~\bibnamefont {Nuss}}, \bibinfo {author}
  {\bibfnamefont {R.~K.}\ \bibnamefont {Kremer}}, \bibinfo {author}
  {\bibfnamefont {R.}~\bibnamefont {Dinnebier}}, \bibinfo {author}
  {\bibfnamefont {M.}~\bibnamefont {Blankenhorn}}, \bibinfo {author}
  {\bibfnamefont {M.}~\bibnamefont {Pakdaman}}, \bibinfo {author}
  {\bibfnamefont {Y.}~\bibnamefont {Matsumoto}}, \bibinfo {author}
  {\bibfnamefont {T.}~\bibnamefont {Takayama}}, \bibinfo {author}
  {\bibfnamefont {K.}~\bibnamefont {Kitagawa}}, \emph {et~al.},\ }\href@noop {}
  {\bibfield  {journal} {\bibinfo  {journal} {Inorg. Chem.}\ }\textbf {\bibinfo
  {volume} {58}},\ \bibinfo {pages} {12888} (\bibinfo {year}
  {2019})}\BibitemShut {NoStop}%
\bibitem [{\citenamefont {Ruck}\ and\ \citenamefont
  {Simon}(1993)}]{ruck1993gd}%
  \BibitemOpen
  \bibfield  {author} {\bibinfo {author} {\bibfnamefont {M.}~\bibnamefont
  {Ruck}}\ and\ \bibinfo {author} {\bibfnamefont {A.}~\bibnamefont {Simon}},\
  }\href@noop {} {\bibfield  {journal} {\bibinfo  {journal} {Z. anorg. allg.
  Chem.}\ }\textbf {\bibinfo {volume} {619}},\ \bibinfo {pages} {327} (\bibinfo
  {year} {1993})}\BibitemShut {NoStop}%
\bibitem [{\citenamefont {Bader}(1990)}]{baderchargeanalysis}%
  \BibitemOpen
  \bibfield  {author} {\bibinfo {author} {\bibfnamefont {R.~F.~W.}\
  \bibnamefont {Bader}},\ }\href@noop {} {\emph {\bibinfo {title} {Atoms in
  Molecules -- A Quantum Theory}}}\ (\bibinfo  {publisher} {Oxford University
  Press: Oxford, U.K.},\ \bibinfo {year} {1990})\BibitemShut {NoStop}%
\bibitem [{\citenamefont {Plumb}\ \emph {et~al.}(2014)\citenamefont {Plumb},
  \citenamefont {Clancy}, \citenamefont {Sandilands}, \citenamefont {Shankar},
  \citenamefont {Hu}, \citenamefont {Burch}, \citenamefont {Kee},\ and\
  \citenamefont {Kim}}]{plumb2014alpha}%
  \BibitemOpen
  \bibfield  {author} {\bibinfo {author} {\bibfnamefont {K.}~\bibnamefont
  {Plumb}}, \bibinfo {author} {\bibfnamefont {J.}~\bibnamefont {Clancy}},
  \bibinfo {author} {\bibfnamefont {L.}~\bibnamefont {Sandilands}}, \bibinfo
  {author} {\bibfnamefont {V.~V.}\ \bibnamefont {Shankar}}, \bibinfo {author}
  {\bibfnamefont {Y.}~\bibnamefont {Hu}}, \bibinfo {author} {\bibfnamefont
  {K.}~\bibnamefont {Burch}}, \bibinfo {author} {\bibfnamefont {H.-Y.}\
  \bibnamefont {Kee}},\ and\ \bibinfo {author} {\bibfnamefont {Y.-J.}\
  \bibnamefont {Kim}},\ }\href@noop {} {\bibfield  {journal} {\bibinfo
  {journal} {Phys. Rev. B}\ }\textbf {\bibinfo {volume} {90}},\ \bibinfo
  {pages} {041112} (\bibinfo {year} {2014})}\BibitemShut {NoStop}%
\bibitem [{\citenamefont {Miura}\ \emph {et~al.}(2007)\citenamefont {Miura},
  \citenamefont {Yasui}, \citenamefont {Sato}, \citenamefont {Igawa},\ and\
  \citenamefont {Kakurai}}]{miura2007new}%
  \BibitemOpen
  \bibfield  {author} {\bibinfo {author} {\bibfnamefont {Y.}~\bibnamefont
  {Miura}}, \bibinfo {author} {\bibfnamefont {Y.}~\bibnamefont {Yasui}},
  \bibinfo {author} {\bibfnamefont {M.}~\bibnamefont {Sato}}, \bibinfo {author}
  {\bibfnamefont {N.}~\bibnamefont {Igawa}},\ and\ \bibinfo {author}
  {\bibfnamefont {K.}~\bibnamefont {Kakurai}},\ }\href@noop {} {\bibfield
  {journal} {\bibinfo  {journal} {J. Phys. Soc. Jpn.}\ }\textbf {\bibinfo
  {volume} {76}},\ \bibinfo {pages} {033705} (\bibinfo {year}
  {2007})}\BibitemShut {NoStop}%
\bibitem [{\citenamefont {Hiley}\ \emph {et~al.}(2014)\citenamefont {Hiley},
  \citenamefont {Lees}, \citenamefont {Fisher}, \citenamefont {Thompsett},
  \citenamefont {Agrestini}, \citenamefont {Smith},\ and\ \citenamefont
  {Walton}}]{hiley2014ruthenium}%
  \BibitemOpen
  \bibfield  {author} {\bibinfo {author} {\bibfnamefont {C.~I.}\ \bibnamefont
  {Hiley}}, \bibinfo {author} {\bibfnamefont {M.~R.}\ \bibnamefont {Lees}},
  \bibinfo {author} {\bibfnamefont {J.~M.}\ \bibnamefont {Fisher}}, \bibinfo
  {author} {\bibfnamefont {D.}~\bibnamefont {Thompsett}}, \bibinfo {author}
  {\bibfnamefont {S.}~\bibnamefont {Agrestini}}, \bibinfo {author}
  {\bibfnamefont {R.~I.}\ \bibnamefont {Smith}},\ and\ \bibinfo {author}
  {\bibfnamefont {R.~I.}\ \bibnamefont {Walton}},\ }\href@noop {} {\bibfield
  {journal} {\bibinfo  {journal} {Angew. Chem.}\ }\textbf {\bibinfo {volume}
  {126}},\ \bibinfo {pages} {4512} (\bibinfo {year} {2014})}\BibitemShut
  {NoStop}%
\bibitem [{\citenamefont {Park}\ \emph {et~al.}(1997)\citenamefont {Park},
  \citenamefont {Martin},\ and\ \citenamefont {Corbett}}]{park1997three}%
  \BibitemOpen
  \bibfield  {author} {\bibinfo {author} {\bibfnamefont {Y.}~\bibnamefont
  {Park}}, \bibinfo {author} {\bibfnamefont {J.~D.}\ \bibnamefont {Martin}},\
  and\ \bibinfo {author} {\bibfnamefont {J.~D.}\ \bibnamefont {Corbett}},\
  }\href@noop {} {\bibfield  {journal} {\bibinfo  {journal} {J. Solid State
  Chem.}\ }\textbf {\bibinfo {volume} {129}},\ \bibinfo {pages} {277} (\bibinfo
  {year} {1997})}\BibitemShut {NoStop}%
\bibitem [{\citenamefont {Momma}\ and\ \citenamefont
  {Izumi}(2011)}]{momma2011vesta}%
  \BibitemOpen
  \bibfield  {author} {\bibinfo {author} {\bibfnamefont {K.}~\bibnamefont
  {Momma}}\ and\ \bibinfo {author} {\bibfnamefont {F.}~\bibnamefont {Izumi}},\
  }\href@noop {} {\bibfield  {journal} {\bibinfo  {journal} {J. Appl.
  Crystallogr.}\ }\textbf {\bibinfo {volume} {44}},\ \bibinfo {pages} {1272}
  (\bibinfo {year} {2011})}\BibitemShut {NoStop}%
\bibitem [{Sup()}]{Supple}%
  \BibitemOpen
  \href@noop {} {\bibinfo  {journal} {See Supplemental Material at [URL] for
  SEM and optical images of the single crystal and calculated band structures}\
  }\BibitemShut {NoStop}%
\bibitem [{\citenamefont
  {Rodr{\'\i}guez-Carvajal}(1993)}]{rodriguez1993recent}%
  \BibitemOpen
\bibfield  {journal} {  }\bibfield  {author} {\bibinfo {author} {\bibfnamefont
  {J.}~\bibnamefont {Rodr{\'\i}guez-Carvajal}},\ }\href@noop {} {\bibfield
  {journal} {\bibinfo  {journal} {Physica B: Condens. Matter}\ }\textbf
  {\bibinfo {volume} {192}},\ \bibinfo {pages} {55} (\bibinfo {year}
  {1993})}\BibitemShut {NoStop}%
\bibitem [{\citenamefont {Giannozzi}\ \emph {et~al.}(2017)\citenamefont
  {Giannozzi}, \citenamefont {Andreussi}, \citenamefont {Brumme}, \citenamefont
  {Bunau}, \citenamefont {Nardelli}, \citenamefont {Calandra}, \citenamefont
  {Car}, \citenamefont {Cavazzoni}, \citenamefont {Ceresoli}, \citenamefont
  {Cococcioni} \emph {et~al.}}]{giannozzi2017advanced}%
  \BibitemOpen
  \bibfield  {author} {\bibinfo {author} {\bibfnamefont {P.}~\bibnamefont
  {Giannozzi}}, \bibinfo {author} {\bibfnamefont {O.}~\bibnamefont
  {Andreussi}}, \bibinfo {author} {\bibfnamefont {T.}~\bibnamefont {Brumme}},
  \bibinfo {author} {\bibfnamefont {O.}~\bibnamefont {Bunau}}, \bibinfo
  {author} {\bibfnamefont {M.~B.}\ \bibnamefont {Nardelli}}, \bibinfo {author}
  {\bibfnamefont {M.}~\bibnamefont {Calandra}}, \bibinfo {author}
  {\bibfnamefont {R.}~\bibnamefont {Car}}, \bibinfo {author} {\bibfnamefont
  {C.}~\bibnamefont {Cavazzoni}}, \bibinfo {author} {\bibfnamefont
  {D.}~\bibnamefont {Ceresoli}}, \bibinfo {author} {\bibfnamefont
  {M.}~\bibnamefont {Cococcioni}}, \emph {et~al.},\ }\href@noop {} {\bibfield
  {journal} {\bibinfo  {journal} {J. Phys. Condens. Matter}\ }\textbf {\bibinfo
  {volume} {29}},\ \bibinfo {pages} {465901} (\bibinfo {year}
  {2017})}\BibitemShut {NoStop}%
\bibitem [{\citenamefont {Perdew}\ \emph {et~al.}(1996)\citenamefont {Perdew},
  \citenamefont {Burke},\ and\ \citenamefont
  {Ernzerhof}}]{perdew1996generalized}%
  \BibitemOpen
  \bibfield  {author} {\bibinfo {author} {\bibfnamefont {J.~P.}\ \bibnamefont
  {Perdew}}, \bibinfo {author} {\bibfnamefont {K.}~\bibnamefont {Burke}},\ and\
  \bibinfo {author} {\bibfnamefont {M.}~\bibnamefont {Ernzerhof}},\ }\href@noop
  {} {\bibfield  {journal} {\bibinfo  {journal} {Phys. Rev. Lett.}\ }\textbf
  {\bibinfo {volume} {77}},\ \bibinfo {pages} {3865} (\bibinfo {year}
  {1996})}\BibitemShut {NoStop}%
\bibitem [{\citenamefont {Scherpelz}\ \emph {et~al.}(2016)\citenamefont
  {Scherpelz}, \citenamefont {Govoni}, \citenamefont {Hamada},\ and\
  \citenamefont {Galli}}]{scherpelz2016implementation}%
  \BibitemOpen
  \bibfield  {author} {\bibinfo {author} {\bibfnamefont {P.}~\bibnamefont
  {Scherpelz}}, \bibinfo {author} {\bibfnamefont {M.}~\bibnamefont {Govoni}},
  \bibinfo {author} {\bibfnamefont {I.}~\bibnamefont {Hamada}},\ and\ \bibinfo
  {author} {\bibfnamefont {G.}~\bibnamefont {Galli}},\ }\href@noop {}
  {\bibfield  {journal} {\bibinfo  {journal} {J. Chem. Theory Comput.}\
  }\textbf {\bibinfo {volume} {12}},\ \bibinfo {pages} {3523} (\bibinfo {year}
  {2016})}\BibitemShut {NoStop}%
\bibitem [{\citenamefont {Kawamura}\ \emph {et~al.}(2014)\citenamefont
  {Kawamura}, \citenamefont {Gohda},\ and\ \citenamefont
  {Tsuneyuki}}]{kawamura2014improved}%
  \BibitemOpen
  \bibfield  {author} {\bibinfo {author} {\bibfnamefont {M.}~\bibnamefont
  {Kawamura}}, \bibinfo {author} {\bibfnamefont {Y.}~\bibnamefont {Gohda}},\
  and\ \bibinfo {author} {\bibfnamefont {S.}~\bibnamefont {Tsuneyuki}},\
  }\href@noop {} {\bibfield  {journal} {\bibinfo  {journal} {Phys. Rev. B}\
  }\textbf {\bibinfo {volume} {89}},\ \bibinfo {pages} {094515} (\bibinfo
  {year} {2014})}\BibitemShut {NoStop}%
\bibitem [{\citenamefont {Tang}\ \emph {et~al.}(2009)\citenamefont {Tang},
  \citenamefont {Sanville},\ and\ \citenamefont {Henkelman}}]{tang2009grid}%
  \BibitemOpen
  \bibfield  {author} {\bibinfo {author} {\bibfnamefont {W.}~\bibnamefont
  {Tang}}, \bibinfo {author} {\bibfnamefont {E.}~\bibnamefont {Sanville}},\
  and\ \bibinfo {author} {\bibfnamefont {G.}~\bibnamefont {Henkelman}},\
  }\href@noop {} {\bibfield  {journal} {\bibinfo  {journal} {J. Phys. Condens.
  Matter}\ }\textbf {\bibinfo {volume} {21}},\ \bibinfo {pages} {084204}
  (\bibinfo {year} {2009})}\BibitemShut {NoStop}%
\bibitem [{\citenamefont {Bl{\"o}chl}(1994)}]{blochl1994projector}%
  \BibitemOpen
  \bibfield  {author} {\bibinfo {author} {\bibfnamefont {P.~E.}\ \bibnamefont
  {Bl{\"o}chl}},\ }\href@noop {} {\bibfield  {journal} {\bibinfo  {journal}
  {Phys. Rev. B}\ }\textbf {\bibinfo {volume} {50}},\ \bibinfo {pages} {17953}
  (\bibinfo {year} {1994})}\BibitemShut {NoStop}%
\bibitem [{\citenamefont {Dal~Corso}(2014)}]{dal2014pseudopotentials}%
  \BibitemOpen
  \bibfield  {author} {\bibinfo {author} {\bibfnamefont {A.}~\bibnamefont
  {Dal~Corso}},\ }\href@noop {} {\bibfield  {journal} {\bibinfo  {journal}
  {Comput. Mater. Sci.}\ }\textbf {\bibinfo {volume} {95}},\ \bibinfo {pages}
  {337} (\bibinfo {year} {2014})}\BibitemShut {NoStop}%
\bibitem [{\citenamefont {M{\"u}ller}\ \emph {et~al.}(1987)\citenamefont
  {M{\"u}ller}, \citenamefont {Takashige},\ and\ \citenamefont
  {Bednorz}}]{muller1987flux}%
  \BibitemOpen
  \bibfield  {author} {\bibinfo {author} {\bibfnamefont {K.}~\bibnamefont
  {M{\"u}ller}}, \bibinfo {author} {\bibfnamefont {M.}~\bibnamefont
  {Takashige}},\ and\ \bibinfo {author} {\bibfnamefont {J.}~\bibnamefont
  {Bednorz}},\ }\href@noop {} {\bibfield  {journal} {\bibinfo  {journal} {Phys.
  Rev. Lett.}\ }\textbf {\bibinfo {volume} {58}},\ \bibinfo {pages} {1143}
  (\bibinfo {year} {1987})}\BibitemShut {NoStop}%
\bibitem [{\citenamefont {Yamanaka}\ \emph {et~al.}(1998)\citenamefont
  {Yamanaka}, \citenamefont {Hotehama},\ and\ \citenamefont
  {Kawaji}}]{yamanaka1998superconductivity}%
  \BibitemOpen
  \bibfield  {author} {\bibinfo {author} {\bibfnamefont {S.}~\bibnamefont
  {Yamanaka}}, \bibinfo {author} {\bibfnamefont {K.-i.}\ \bibnamefont
  {Hotehama}},\ and\ \bibinfo {author} {\bibfnamefont {H.}~\bibnamefont
  {Kawaji}},\ }\href@noop {} {\bibfield  {journal} {\bibinfo  {journal}
  {Nature}\ }\textbf {\bibinfo {volume} {392}},\ \bibinfo {pages} {580}
  (\bibinfo {year} {1998})}\BibitemShut {NoStop}%
\bibitem [{\citenamefont {Hotehama}\ \emph {et~al.}(2009)\citenamefont
  {Hotehama}, \citenamefont {Koiwasaki}, \citenamefont {Umemoto}, \citenamefont
  {Yamanaka},\ and\ \citenamefont {Tou}}]{hotehama2009effect}%
  \BibitemOpen
  \bibfield  {author} {\bibinfo {author} {\bibfnamefont {K.-i.}\ \bibnamefont
  {Hotehama}}, \bibinfo {author} {\bibfnamefont {T.}~\bibnamefont {Koiwasaki}},
  \bibinfo {author} {\bibfnamefont {K.}~\bibnamefont {Umemoto}}, \bibinfo
  {author} {\bibfnamefont {S.}~\bibnamefont {Yamanaka}},\ and\ \bibinfo
  {author} {\bibfnamefont {H.}~\bibnamefont {Tou}},\ }\href@noop {} {\bibfield
  {journal} {\bibinfo  {journal} {J. Phys. Soc. Jpn.}\ }\textbf {\bibinfo
  {volume} {79}},\ \bibinfo {pages} {014707} (\bibinfo {year}
  {2009})}\BibitemShut {NoStop}%
\bibitem [{\citenamefont {Tsuchiya}\ \emph {et~al.}(2012)\citenamefont
  {Tsuchiya}, \citenamefont {Yamada}, \citenamefont {Tanda}, \citenamefont
  {Ichimura}, \citenamefont {Terashima}, \citenamefont {Kurita}, \citenamefont
  {Kodama},\ and\ \citenamefont {Uji}}]{tsuchiya2012fluctuating}%
  \BibitemOpen
  \bibfield  {author} {\bibinfo {author} {\bibfnamefont {S.}~\bibnamefont
  {Tsuchiya}}, \bibinfo {author} {\bibfnamefont {J.-i.}\ \bibnamefont
  {Yamada}}, \bibinfo {author} {\bibfnamefont {S.}~\bibnamefont {Tanda}},
  \bibinfo {author} {\bibfnamefont {K.}~\bibnamefont {Ichimura}}, \bibinfo
  {author} {\bibfnamefont {T.}~\bibnamefont {Terashima}}, \bibinfo {author}
  {\bibfnamefont {N.}~\bibnamefont {Kurita}}, \bibinfo {author} {\bibfnamefont
  {K.}~\bibnamefont {Kodama}},\ and\ \bibinfo {author} {\bibfnamefont
  {S.}~\bibnamefont {Uji}},\ }\href@noop {} {\bibfield  {journal} {\bibinfo
  {journal} {Phys. Rev. B}\ }\textbf {\bibinfo {volume} {85}},\ \bibinfo
  {pages} {220506} (\bibinfo {year} {2012})}\BibitemShut {NoStop}%
\bibitem [{\citenamefont {Werthamer}\ \emph {et~al.}(1966)\citenamefont
  {Werthamer}, \citenamefont {Helfand},\ and\ \citenamefont
  {Hohenberg}}]{werthamer1966temperature}%
  \BibitemOpen
  \bibfield  {author} {\bibinfo {author} {\bibfnamefont {N.}~\bibnamefont
  {Werthamer}}, \bibinfo {author} {\bibfnamefont {E.}~\bibnamefont {Helfand}},\
  and\ \bibinfo {author} {\bibfnamefont {P.}~\bibnamefont {Hohenberg}},\
  }\href@noop {} {\bibfield  {journal} {\bibinfo  {journal} {Phy. Rev.}\
  }\textbf {\bibinfo {volume} {147}},\ \bibinfo {pages} {295} (\bibinfo {year}
  {1966})}\BibitemShut {NoStop}%
\bibitem [{\citenamefont {Henn}\ \emph {et~al.}(1996)\citenamefont {Henn},
  \citenamefont {Schnelle}, \citenamefont {Kremer},\ and\ \citenamefont
  {Simon}}]{henn1996bulk}%
  \BibitemOpen
  \bibfield  {author} {\bibinfo {author} {\bibfnamefont {R.}~\bibnamefont
  {Henn}}, \bibinfo {author} {\bibfnamefont {W.}~\bibnamefont {Schnelle}},
  \bibinfo {author} {\bibfnamefont {R.}~\bibnamefont {Kremer}},\ and\ \bibinfo
  {author} {\bibfnamefont {A.}~\bibnamefont {Simon}},\ }\href@noop {}
  {\bibfield  {journal} {\bibinfo  {journal} {Phys. Rev. Lett.}\ }\textbf
  {\bibinfo {volume} {77}},\ \bibinfo {pages} {374} (\bibinfo {year}
  {1996})}\BibitemShut {NoStop}%
\bibitem [{\citenamefont {Ahn}\ \emph {et~al.}(2000)\citenamefont {Ahn},
  \citenamefont {Kremer},\ and\ \citenamefont
  {Simon}}]{ahn2000superconductivity}%
  \BibitemOpen
  \bibfield  {author} {\bibinfo {author} {\bibfnamefont {K.}~\bibnamefont
  {Ahn}}, \bibinfo {author} {\bibfnamefont {R.}~\bibnamefont {Kremer}},\ and\
  \bibinfo {author} {\bibfnamefont {A.}~\bibnamefont {Simon}},\ }\href@noop {}
  {\bibfield  {journal} {\bibinfo  {journal} {J. Alloys Compd.}\ }\textbf
  {\bibinfo {volume} {303}},\ \bibinfo {pages} {257} (\bibinfo {year}
  {2000})}\BibitemShut {NoStop}%
\bibitem [{\citenamefont {Ryazanov}\ \emph {et~al.}(2006)\citenamefont
  {Ryazanov}, \citenamefont {Simon},\ and\ \citenamefont
  {Mattausch}}]{ryazanov2006la2tei2}%
  \BibitemOpen
  \bibfield  {author} {\bibinfo {author} {\bibfnamefont {M.}~\bibnamefont
  {Ryazanov}}, \bibinfo {author} {\bibfnamefont {A.}~\bibnamefont {Simon}},\
  and\ \bibinfo {author} {\bibfnamefont {H.}~\bibnamefont {Mattausch}},\
  }\href@noop {} {\bibfield  {journal} {\bibinfo  {journal} {Inorg. Chem.}\
  }\textbf {\bibinfo {volume} {45}},\ \bibinfo {pages} {10728} (\bibinfo {year}
  {2006})}\BibitemShut {NoStop}%
\bibitem [{\citenamefont {Keimer}\ \emph {et~al.}(2015)\citenamefont {Keimer},
  \citenamefont {Kivelson}, \citenamefont {Norman}, \citenamefont {Uchida},\
  and\ \citenamefont {Zaanen}}]{keimer2015quantum}%
  \BibitemOpen
  \bibfield  {author} {\bibinfo {author} {\bibfnamefont {B.}~\bibnamefont
  {Keimer}}, \bibinfo {author} {\bibfnamefont {S.~A.}\ \bibnamefont
  {Kivelson}}, \bibinfo {author} {\bibfnamefont {M.~R.}\ \bibnamefont
  {Norman}}, \bibinfo {author} {\bibfnamefont {S.}~\bibnamefont {Uchida}},\
  and\ \bibinfo {author} {\bibfnamefont {J.}~\bibnamefont {Zaanen}},\
  }\href@noop {} {\bibfield  {journal} {\bibinfo  {journal} {Nature}\ }\textbf
  {\bibinfo {volume} {518}},\ \bibinfo {pages} {179} (\bibinfo {year}
  {2015})}\BibitemShut {NoStop}%
\bibitem [{\citenamefont {Si}\ \emph {et~al.}(2016)\citenamefont {Si},
  \citenamefont {Yu},\ and\ \citenamefont {Abrahams}}]{si2016high}%
  \BibitemOpen
  \bibfield  {author} {\bibinfo {author} {\bibfnamefont {Q.}~\bibnamefont
  {Si}}, \bibinfo {author} {\bibfnamefont {R.}~\bibnamefont {Yu}},\ and\
  \bibinfo {author} {\bibfnamefont {E.}~\bibnamefont {Abrahams}},\ }\href@noop
  {} {\bibfield  {journal} {\bibinfo  {journal} {Nat. Rev. Mater.}\ }\textbf
  {\bibinfo {volume} {1}},\ \bibinfo {pages} {1} (\bibinfo {year}
  {2016})}\BibitemShut {NoStop}%
\bibitem [{\citenamefont {Berthier}\ \emph {et~al.}(1976)\citenamefont
  {Berthier}, \citenamefont {Molini{\'e}},\ and\ \citenamefont
  {J{\'e}rome}}]{berthier1976evidence}%
  \BibitemOpen
  \bibfield  {author} {\bibinfo {author} {\bibfnamefont {C.}~\bibnamefont
  {Berthier}}, \bibinfo {author} {\bibfnamefont {P.}~\bibnamefont
  {Molini{\'e}}},\ and\ \bibinfo {author} {\bibfnamefont {D.}~\bibnamefont
  {J{\'e}rome}},\ }\href@noop {} {\bibfield  {journal} {\bibinfo  {journal}
  {Solid State Commun.}\ }\textbf {\bibinfo {volume} {18}},\ \bibinfo {pages}
  {1393} (\bibinfo {year} {1976})}\BibitemShut {NoStop}%
\bibitem [{\citenamefont {Morosan}\ \emph {et~al.}(2006)\citenamefont
  {Morosan}, \citenamefont {Zandbergen}, \citenamefont {Dennis}, \citenamefont
  {Bos}, \citenamefont {Onose}, \citenamefont {Klimczuk}, \citenamefont
  {Ramirez}, \citenamefont {Ong},\ and\ \citenamefont
  {Cava}}]{morosan2006superconductivity}%
  \BibitemOpen
  \bibfield  {author} {\bibinfo {author} {\bibfnamefont {E.}~\bibnamefont
  {Morosan}}, \bibinfo {author} {\bibfnamefont {H.~W.}\ \bibnamefont
  {Zandbergen}}, \bibinfo {author} {\bibfnamefont {B.}~\bibnamefont {Dennis}},
  \bibinfo {author} {\bibfnamefont {J.}~\bibnamefont {Bos}}, \bibinfo {author}
  {\bibfnamefont {Y.}~\bibnamefont {Onose}}, \bibinfo {author} {\bibfnamefont
  {T.}~\bibnamefont {Klimczuk}}, \bibinfo {author} {\bibfnamefont
  {A.}~\bibnamefont {Ramirez}}, \bibinfo {author} {\bibfnamefont
  {N.}~\bibnamefont {Ong}},\ and\ \bibinfo {author} {\bibfnamefont {R.~J.}\
  \bibnamefont {Cava}},\ }\href@noop {} {\bibfield  {journal} {\bibinfo
  {journal} {Nat. Phys.}\ }\textbf {\bibinfo {volume} {2}},\ \bibinfo {pages}
  {544} (\bibinfo {year} {2006})}\BibitemShut {NoStop}%
\bibitem [{\citenamefont {Chen}\ \emph {et~al.}(2021)\citenamefont {Chen},
  \citenamefont {Wang}, \citenamefont {Yin}, \citenamefont {Gu}, \citenamefont
  {Jiang}, \citenamefont {Tu}, \citenamefont {Gong}, \citenamefont {Uwatoko},
  \citenamefont {Sun}, \citenamefont {Lei} \emph {et~al.}}]{chen2021double}%
  \BibitemOpen
  \bibfield  {author} {\bibinfo {author} {\bibfnamefont {K.}~\bibnamefont
  {Chen}}, \bibinfo {author} {\bibfnamefont {N.}~\bibnamefont {Wang}}, \bibinfo
  {author} {\bibfnamefont {Q.}~\bibnamefont {Yin}}, \bibinfo {author}
  {\bibfnamefont {Y.}~\bibnamefont {Gu}}, \bibinfo {author} {\bibfnamefont
  {K.}~\bibnamefont {Jiang}}, \bibinfo {author} {\bibfnamefont
  {Z.}~\bibnamefont {Tu}}, \bibinfo {author} {\bibfnamefont {C.}~\bibnamefont
  {Gong}}, \bibinfo {author} {\bibfnamefont {Y.}~\bibnamefont {Uwatoko}},
  \bibinfo {author} {\bibfnamefont {J.}~\bibnamefont {Sun}}, \bibinfo {author}
  {\bibfnamefont {H.}~\bibnamefont {Lei}}, \emph {et~al.},\ }\href@noop {}
  {\bibfield  {journal} {\bibinfo  {journal} {Phys. Rev. Lett.}\ }\textbf
  {\bibinfo {volume} {126}},\ \bibinfo {pages} {247001} (\bibinfo {year}
  {2021})}\BibitemShut {NoStop}%
\bibitem [{\citenamefont {C.~Kobayashi}\ \emph {et~al.}(2011)\citenamefont
  {C.~Kobayashi}, \citenamefont {Irie}, \citenamefont {Yamaura}, \citenamefont
  {Hiroi},\ and\ \citenamefont {Murata}}]{c2011superconductivity}%
  \BibitemOpen
  \bibfield  {author} {\bibinfo {author} {\bibfnamefont {T.}~\bibnamefont
  {C.~Kobayashi}}, \bibinfo {author} {\bibfnamefont {Y.}~\bibnamefont {Irie}},
  \bibinfo {author} {\bibfnamefont {J.-i.}\ \bibnamefont {Yamaura}}, \bibinfo
  {author} {\bibfnamefont {Z.}~\bibnamefont {Hiroi}},\ and\ \bibinfo {author}
  {\bibfnamefont {K.}~\bibnamefont {Murata}},\ }\href@noop {} {\bibfield
  {journal} {\bibinfo  {journal} {J. Phys. Soc. Jpn.}\ }\textbf {\bibinfo
  {volume} {80}},\ \bibinfo {pages} {023715} (\bibinfo {year}
  {2011})}\BibitemShut {NoStop}%
\bibitem [{\citenamefont {Zehetmayer}(2013)}]{zehetmayer2013review}%
  \BibitemOpen
  \bibfield  {author} {\bibinfo {author} {\bibfnamefont {M.}~\bibnamefont
  {Zehetmayer}},\ }\href@noop {} {\bibfield  {journal} {\bibinfo  {journal}
  {Supercond. Sci. Technol.}\ }\textbf {\bibinfo {volume} {26}},\ \bibinfo
  {pages} {043001} (\bibinfo {year} {2013})}\BibitemShut {NoStop}%
\bibitem [{\citenamefont {Mazin}\ and\ \citenamefont
  {Antropov}(2003)}]{mazin2003electronic}%
  \BibitemOpen
  \bibfield  {author} {\bibinfo {author} {\bibfnamefont {I.}~\bibnamefont
  {Mazin}}\ and\ \bibinfo {author} {\bibfnamefont {V.}~\bibnamefont
  {Antropov}},\ }\href@noop {} {\bibfield  {journal} {\bibinfo  {journal}
  {Phys. C: Supercond.}\ }\textbf {\bibinfo {volume} {385}},\ \bibinfo {pages}
  {49} (\bibinfo {year} {2003})}\BibitemShut {NoStop}%
\bibitem [{\citenamefont {Petrovi{\'c}}\ \emph {et~al.}(2011)\citenamefont
  {Petrovi{\'c}}, \citenamefont {Lortz}, \citenamefont {Santi}, \citenamefont
  {Berthod}, \citenamefont {Dubois}, \citenamefont {Decroux}, \citenamefont
  {Demuer}, \citenamefont {Antunes}, \citenamefont {Par{\'e}}, \citenamefont
  {Salloum} \emph {et~al.}}]{petrovic2011multiband}%
  \BibitemOpen
  \bibfield  {author} {\bibinfo {author} {\bibfnamefont {A.~P.}\ \bibnamefont
  {Petrovi{\'c}}}, \bibinfo {author} {\bibfnamefont {R.}~\bibnamefont {Lortz}},
  \bibinfo {author} {\bibfnamefont {G.}~\bibnamefont {Santi}}, \bibinfo
  {author} {\bibfnamefont {C.}~\bibnamefont {Berthod}}, \bibinfo {author}
  {\bibfnamefont {C.}~\bibnamefont {Dubois}}, \bibinfo {author} {\bibfnamefont
  {M.}~\bibnamefont {Decroux}}, \bibinfo {author} {\bibfnamefont
  {A.}~\bibnamefont {Demuer}}, \bibinfo {author} {\bibfnamefont {A.~B.}\
  \bibnamefont {Antunes}}, \bibinfo {author} {\bibfnamefont {A.}~\bibnamefont
  {Par{\'e}}}, \bibinfo {author} {\bibfnamefont {D.}~\bibnamefont {Salloum}},
  \emph {et~al.},\ }\href@noop {} {\bibfield  {journal} {\bibinfo  {journal}
  {Phys. Rev. Lett.}\ }\textbf {\bibinfo {volume} {106}},\ \bibinfo {pages}
  {017003} (\bibinfo {year} {2011})}\BibitemShut {NoStop}%
\end{thebibliography}%

\end{document}